\documentclass{article}
\usepackage{amsmath,accents}
\usepackage{url}
\usepackage{eufrak}
\usepackage{amsfonts}
\usepackage{amssymb}
\usepackage{amsthm}
\usepackage{comment}
\usepackage[figtopcap]{subfigure}
\usepackage[colorlinks,citecolor=blue,urlcolor=blue,linkcolor=blue]{hyperref}
\usepackage[dvipsnames]{xcolor}
\definecolor{cpublication}{rgb}{0.858, 0.4, 0.1}
\definecolor{carxiv}{rgb}{0.858, 0.188, 0.478}

\newcommand\be{\begin{equation}}
\newcommand\ee{\end{equation}}
\newcommand\bea{\begin{eqnarray}}
\newcommand\eea{\end{eqnarray}}
\newcommand\bsp{\begin{split}}
\newcommand\esp{\end{split}}

\usepackage[sort&compress,numbers]{natbib}
\usepackage{tikz}
\usepackage{tikzsymbols}
\usepackage[utf8]{inputenc} 
\usepackage{times}
\usepackage{mathtools}
\usepackage{footnote}
\makesavenoteenv{table}

\title{\bf The Hamiltonian constraint in the symmetric teleparallel equivalent of general relativity}
\author{Mar\'ia-Jos\'e Guzm\'an\\ \\
\it Laboratory of Theoretical Physics, Institute of Physics, University of Tartu,\\ \it W. Ostwaldi 1, 50411 Tartu, Estonia\\ \\
\small mjguzman@ut.ee}

\date{\today}

\begin{document}

\maketitle

\abstract{General relativity (GR) admits two alternative formulations with the same dynamics attributing the gravitational phenomena to torsion or nonmetricity of the manifold's connection. They lead, respectively, to the teleparallel equivalent of general relativity (TEGR) and the symmetric teleparallel equivalent of general relativity (STEGR). In this work, we focus on STEGR and present its differences with the conventional, curvature-based GR. We exhibit the 3+1 decomposition of the STEGR Lagrangian in the coincident gauge and present the Hamiltonian, the Hamiltonian and momenta constraints, and Hamilton's equations. For a particular case of spherical symmetry, we explicitly show the differences in the Hamiltonian and the Hamiltonian constraint between GR and STEGR. We finally discuss the implications that these differences, which represent genuine different features between the two formulations of gravity, might encompass to numerical relativity.}

\section{Introduction}

The observational detection of gravitational waves has unveiled an opportunity to explore the farthest uncharted reaches of gravity, where the most violent events, such as binary black hole mergers, offer insights into physics beyond general relativity (GR) \cite{LIGOScientific:2016aoc,LISA:2022kgy}. With this recent breakthrough it becomes imperative to push further advancements in numerical relativity simulations, since a detailed understanding of the nonlinear dynamics of strong gravitational fields necessitates to solve Einstein's equations numerically, as perturbative approaches are doomed to fail in this regime. 

General relativity, the founding stone of our current understanding of gravitational phenomena and also the base of numerical relativity as currently known, has been challenged by Einstein himself through the teleparallel framework in an attempt to find a unified theory for gravity and electromagnetism \cite{Aldrovandi2013,Unzicker}. Although this teleparallel episode ended up with the realization that the extra degrees of freedom introduced by the \textit{vierbein} or tetrad field correspond to Lorentz transformations instead of the components of the electromagnetic field, it served as a branch of research that had its ups and downs, with a recent surge of interest. The teleparallel framework has inspired the proposition of the so-called geometric trinity of gravity \cite{Jimenez19}. This is fundamented in the fact that GR presupposes that the starting point for the variational principle leading to Einstein's field equations is the Ricci scalar $\mathbb R$ with the Levi-Civita connection depending purely on the metric tensor. It turns out that the same equations of motion can be obtained from considering the torsion or nonmetricity scalars $\mathbb T$ or $\mathbb Q$, respectively \cite{Jimenez19}. The majority of the body of the literature has been devoted to metric teleparallel gravity and its modifications \cite{Bahamonde:2021gfp} with a recent surge of interest in symmetric teleparallel gravity \cite{Heisenberg:2023lru}. 

In a regular setup such as general relativity written in the tetrad formalism, the tetrad encodes exactly the same number of degrees of freedom as the metric tensor, and all the additional functions coming from the nonsymmetric-matrix representation of the tetrad components are removed by local Lorentz transformations. Meanwhile, the symmetric teleparallel framework is written in terms of the nonmetricity tensor, a rank-3 tensor symmetric in its last two indices. The Lagrangians of TEGR and STEGR differ from the general relativity Lagrangian by a boundary term, harmless for the dynamics but relevant for physics depending on the boundary terms. In this regard, it is asserted that physically relevant differences may appear when considering Noether currents in gravitating systems, since they are sensitive to the boundary terms in the action. \cite{BeltranJimenez:2021kpj,Gomes:2022vrc}. The action functional is key in the computation of the Noether currents corresponding to the symmetries of the theory, and in particular it is used to compute conserved charges such as the energy and the entropy \cite{Emtsova:2022uij,Jarv:2023anw}. The action is also crucial for path integral approaches to quantization of gravity. Finally, the resemblance of teleparallel gravities to a Yang-Mills theory  has also been noted as an advantage of the teleparallel approach, making it potentially more suitable as a starting point for canonical quantization \cite{Aldrovandi2013}. 

In addition, there is a scarcely explored aspect concerning to how the boundary terms alter the canonical momenta, the Hamiltonian, and therefore the canonical structure. In particular, such a change in the Hamiltonian does not alter the equations of motion but can affect the Hamiltonian constraint and Hamilton's equations, which play a role in the numerical relativity setup. Since in TEGR and STEGR the boundary terms are different and can even have multiple ways of being written \cite{Boehmer:2021aji}, there are numerous and unknown different ways of reformulating the constraints, offering a whole area of research to follow. There have been significatively more efforts in the analysis of the Hamiltonian structure of TEGR \cite{Maluf:1994ji,Maluf:1998ae,Blagojevic:2000qs,Maluf:2001rg,daRochaNeto:2010xls,Maluf:2013gaa,Okolow:2011nq,Okolow:2013lwa,Ferraro:2016wht,Blixt:2018znp,Guzman:2020kgh,Blixt:2020ekl,Golovnev:2021omn} and its modifications \cite{Li:2011rn,Ferraro:2018tpu,Ferraro:2018axk,Blagojevic:2020dyq,Golovnev:2020nln,Golovnev:2020zpv,Barker:2022kdk}, but recent works in STEGR \cite{DAmbrosio:2020nqu,Dambrosio:2020wbi} and nonlinear extensions \cite{Hu:2022anq,DAmbrosio:2023asf,Tomonari:2023wcs} are also starting to flourish, not necessarily with better outcomes than for its nonlinear TEGR counterpart. Regarding the 3+1 decomposition in the teleparallel framework, we find Ref. \cite{Capozziello:2021pcg} in the Lagrangian formalism and Ref. \cite{Pati:2022nwi} in the Hamiltonian formalism, where in the latter some criticism has been presented regarding the former reference. Recently it has also been presented the 3+1 formulation in metric teleparallel gravity in a first-order formulation in \cite{Peshkov:2022cbi}.

The aim of this paper is to explore the consequences of adopting the symmetric teleparallel equivalent of general relativity as the foundational framework for numerical relativity. We focus on the 3+1 decomposition of the Lagrangian in the STEGR, and derive general expressions for the Hamiltonian, the Hamiltonian constraint, and Hamilton's equations. We also display a simple example of spherical symmetry where all the previous expressions are computed, showing that they are significantly simplified compared with its GR counterpart. The outline of the paper is structured as follows. We briefly introduce the relevant theoretical framework for the teleparallel geometry in Sec. \ref{sec:theo}, and present the gravitational sector built on top of the geometry in Sec. \ref{sec:stegr}. We show the ADM decomposition in the metric, and with it perform the 3+1 split in the STEGR Lagrangian in Sec. \ref{sec:split}. We present Hamilton's equations for STEGR in Sec. \ref{sec:Hamilteqns}. We provide a specific example of spherical symmetry in Sec. \ref{sec:example}. After a short discussion on the future perspectives for our research in Sec. \ref{sec:ext}, we present our conclusions in Sec. \ref{sec:concl}.

\section{The metric affine framework, metric and symmetric teleparallel gravity}
\label{sec:theo}
A general manifold describing spacetime can be represented in terms of a metric tensor $g_{\mu\nu}$ defining the metric structure, and an affine connection $\Gamma^{\rho}{}_{\mu\nu}$ defining the notion of parallel transport along it. These two structures are \textit{a priori} independent, but depending on the subclass of geometry we describe, they might be interdependent. A connection that makes the covariant derivative of the metric zero is said to be a metric connection, and the departure from metricity is measured by the nonmetricity tensor 
\be 
Q_{\rho\mu\nu} = \nabla_{\rho} g_{\mu\nu}.
\label{nonm}
\ee 
The antisymmetric part of the connection defines the torsion tensor \footnote{Our convention will be $A_{[\mu\nu]} = \frac12( A_{\mu\nu} - A_{\nu\mu})$.}
\be 
T^{\rho}{}_{\mu\nu} = 2\Gamma^{\rho}{}_{[\mu\nu]}
\label{tors}
\ee 
General relativity is built in a geometric framework that \textit{assumes} that both nonmetricity and torsion of the linear connection are zero. This choice leads to a unique expression for the connection, the well known Levi-Civita connection, that we will denote with an overcircle and we define as usual
\be 
\overset{\circ}{\Gamma}{}^{\rho}{}_{\mu\nu} = \dfrac12 g^{\rho\lambda}(\partial_{\mu}g_{\nu\lambda} + \partial_{\nu}g_{\mu\lambda}-\partial_{\lambda}g_{\mu\nu}  ).
\ee 
Therefore, the covariant derivative associated to it, denoted by $\overset{\circ}{\nabla}$, satisfies $\overset{\circ}{\nabla}_{\rho}g_{\mu\nu}=0$. 
However, in the most general case, the linear connection is composed of three parts, or in different words, it corresponds to the Levi-Civita connection plus two extra terms, as in the following
\be 
\Gamma^{\rho}{}_{\mu\nu} = \overset{\circ}{\Gamma^{\rho}} {}_{\mu\nu} + K^{\rho}{}_{\mu\nu} + L^{\rho}{}_{\mu\nu},
\label{affcon}
\ee 
where we have defined the contortion
\begin{equation}
    K_{\alpha\mu\nu} = \dfrac12 \left( T_{\mu\alpha\nu} + T_{\nu\alpha\mu} - T_{\alpha\mu\nu} \right)
\end{equation}
and the distortion
\be 
L^{\rho}{}_{\mu\nu} = \dfrac12 ( Q^{\rho}{}_{\mu\nu} - Q_{\mu \ \nu}^{\ \rho} + Q_{\nu \ \mu}^{\ \rho})
\label{distortion}
\ee 
tensors. The Riemann tensor of the most general linear connection is
\be 
{R}{}^{\rho}{}_{\lambda\mu\nu} = \partial_{\mu}\Gamma^{\rho}{}_{\nu\lambda} - \partial_{\nu}\Gamma^{\rho}{}_{\mu\lambda} + \Gamma^{\rho}{}_{\mu\sigma}\Gamma^{\sigma}{}_{\nu\lambda} - \Gamma^{\rho}{}_{\nu\sigma}\Gamma^{\sigma}{}_{\mu\lambda}.
\label{curv}
\ee 
From these definitions it is possible to obtain three geometric frameworks that are relevant to general relativity:
\begin{itemize}
    \item For a connection with vanishing nonmetricity \eqref{nonm} and torsion \eqref{tors}, we recover the Levi-Civita connection.
    \item For a connection vanishing the Riemann tensor \eqref{curv} and the nonmetricity \eqref{nonm}, the torsion is the only remaining geometrical object, and from it can be built the torsion scalar $\mathbb T$ which is equivalent to the Ricci scalar $\mathbb R$ by a boundary term. The connection however, is not unequivocally determined, and can be found to depend on matrices belonging to the Lorentz group. 
    \item For a connection with vanishing Riemann tensor \eqref{curv} and vanishing torsion \eqref{tors}, the nonmetricity tensor is the remaining geometric object that builds the nonmetricity scalar $\mathbb Q$ differing from $\mathbb R$ by a boundary term. The connection is again not unique, and can be written in terms of a set of four scalars $\xi^{a}$ representing a diffeomorphism.
\end{itemize}
In the following we will focus on the third case, and we will delve in more mathematical details and in preparing the Lagrangian for the symmetric teleparallel equivalent of general relativity (STEGR) for the 3+1 formalism.

\section{STEGR}
\label{sec:stegr}

The symmetric teleparallel framework emerges when considering a connection that is \textit{symmetric} in the last two indices (hence its name symmetric), which means that it is a connection with vanishing torsion. Additionally, this geometrical framework imposes the Riemann tensor to vanish. Unlike GR where the torsionless, metric compatible Levi-Civita connection is uniquely determined by the metric and its derivatives, the symmetric teleparallel connection is not unique and in its more general form is written as
\be 
\Gamma^{\rho}{}_{\mu\nu} = \frac{\partial^2 \xi^{a} }{\partial x^{\mu} \partial x^{\nu} } \frac{\partial x^{\rho} }{\partial \xi^{a} }.
\label{coing}
\ee 
The connection components are given by an arbitrary coordinate transformation $\xi^{a}(x^{\nu})$ starting from a frame with zero connection. It is said that we are in the coincident gauge when the coordinate systems $\xi^{a}$ and $x^{\mu}$ coincide, and the connection \eqref{coing} is trivialized and vanishes \cite{BeltranJimenez:2017tkd}. One of the consequences of this choice is that the nonmetricity tensor acquires the simple form $Q_{\rho\mu\nu}=\partial_{\rho}g_{\mu\nu}$. However, in the following we will  consider more convenient to introduce the STEGR Lagrangian for a general connection \eqref{coing} without fixing the gauge yet, then after that we will specialize to the case of the coincident gauge.

Consequently, in the symmetric teleparallel geometric framework, the affine connection \eqref{affcon} becomes
\be 
\Gamma^{\rho}{}_{\mu\nu} = \overset{\circ}{\Gamma^{\rho}} {}_{\mu\nu} + L^{\rho}{}_{\mu\nu},
\label{stegrcon}
\ee 
and if we compute the Ricci tensor of it through the expression \eqref{curv}, we obtain
\be 
R_{\mu\nu} = \overset{\circ}{R}_{\mu\nu} + L^{\rho}{}_{\mu\nu}L^{\sigma}{}_{\sigma\rho} - L^{\rho}{}_{\mu\sigma}L^{\sigma}{}_{\rho\nu} + \overset{\circ}{\nabla}_{\rho}L^{\rho}{}_{\mu\nu} - \overset{\circ}{\nabla}_{\mu}L^{\rho}{}_{\rho\nu}.
\ee 
Contracting this expression with the inverse metric $g^{\mu\nu}$ gives us a relation among scalars
\be 
\mathbb R = \overset{\circ}{\mathbb R} - \mathbb Q + \overset{\circ}{\nabla}_{\mu}(Q^{\mu}-\tilde{Q}^{\mu} ),
\label{nonmeq}
\ee 
where it has been defined the following two (independent) traces of the nonmetricity tensor $$Q^{\mu}= Q^{\mu \nu}_{ \hphantom{\mu\nu} \nu}, \ \ \ \tilde{Q}^{\mu} = Q_{\nu}{}^{\mu\nu}=Q_{\nu}{}^{\nu\mu}.$$ 
Meanwhile, $\overset{\circ}{\mathbb R}$ is the Ricci scalar obtained from the Levi-Civita connection, and $\mathbb Q$ is the so-called nonmetricity scalar and is obtained in terms of the distortion tensor as
\be 
\mathbb Q =  L^{\rho}{}_{\mu\nu}L^{\sigma}{}_{\sigma\rho} - L^{\rho}{}_{\mu\sigma}L^{\sigma}{}_{\rho\nu}.
\ee 
However, in \eqref{nonmeq} we have still not imposed that we would like to work in a teleparallel spacetime. For this to happen, our connection must vanish the full curvature $\mathbb R=0$ too, and in this way it is then obtained the well-known equivalence
\be 
\overset{\circ}{\mathbb R} = \mathbb Q - \overset{\circ}{\nabla}_{\mu}(Q^{\mu} - \tilde{Q}^{\mu})
\label{equiv}
\ee 
which states that the Ricci $\mathbb R$ scalar differs from the nonmetricity scalar $\mathbb Q$ by just a boundary term. With these ingredients we are ready to obtain the action for the symmetric teleparallel equivalent of general relativity. For this, let us consider the Einstein-Hilbert action, and replace the Ricci scalar on it in terms of the nonmetricity scalar $\mathbb Q$ and the boundary term from \eqref{equiv}, obtaining
\be 
\mathcal{S} = \dfrac{1}{2\kappa}\int d^4x \sqrt{-g} \overset{\circ}{R} = \dfrac{1}{2\kappa} \int d^4x \sqrt{-g} \mathbb Q,
\label{Q}
\ee 
if integrating out the boundary term. 

The nonmetricity scalar $\mathbb Q$ can be rewritten in several alternative forms taking \eqref{stegrcon} as a starting point. If we replace the definition of the distortion $L^{\rho}{}_{\mu\nu}$ in terms of the nonmetricity tensor $Q_{\rho\mu\nu}$ as in \eqref{distortion}, it is obtained the more common expression found in the literature
\be 
\mathbb Q = -\dfrac14 Q_{\rho\mu\nu} Q^{\rho\mu\nu} + \dfrac12 Q_{\rho\mu\nu} Q^{\mu\nu\rho} + \dfrac14 Q_{\mu} Q^{\mu}  -\dfrac12 Q_{\mu} \tilde{Q}^{\mu}.
\label{Qstegr}
\ee 
This can also be written as an explicit quadratic combination of the nonmetricity tensor
\be 
\mathbb Q = \dfrac14 ( -g^{\alpha\rho} g^{\beta\mu} g^{\gamma\nu} + 2g^{\alpha\nu} g^{\beta\mu} g^{\gamma\rho} + g^{\alpha\rho} g^{\beta\gamma} g^{\mu\nu} - 2g^{\alpha\beta} g^{\mu\nu} g^{\gamma\rho} ) Q_{\rho\mu\nu} Q_{\alpha\beta\gamma}.
\ee 
Now we would like to specify these expression to the coincident gauge where $Q_{\rho\mu\nu} = \partial_{\rho} g_{\mu\nu}$. From the previous formula it is easy to see that $\mathbb Q$ corresponds to
\be
\mathbb Q = -\dfrac14 (g^{\mu\nu}g^{\rho\lambda}g^{\gamma\delta} + 2g^{\gamma\mu}g^{\delta\rho}g^{\lambda\nu} - g^{\mu\gamma}g^{\nu\delta}g^{\rho\lambda} - 2g^{\mu\nu}g^{\rho\gamma}g^{\delta\lambda} )\partial_{\rho}g_{\mu\nu}\partial_{\lambda}g_{\gamma\delta}.
\label{Qmetric}
\ee 
We can work out the same exercise for the boundary term found in  \eqref{equiv}. For this, it is useful to rewrite the nonmetricity traces in the coincident gauge
\be 
\begin{split}
Q^{\mu} & =Q^{\mu\nu}{}_{\nu} =g^{\alpha\mu}g^{\beta\nu} Q_{\alpha\beta\nu} = g^{\alpha\mu}g^{\beta\nu} \partial_{\alpha}g_{\beta\nu}, \\
\tilde{Q}^{\mu} & =Q^{\nu\mu}{}_{\nu} = g^{\alpha\nu} g^{\beta\mu} Q_{\alpha\beta\nu} = g^{\alpha\nu} g^{\beta\mu} \partial_{\alpha}g_{\beta\nu}.
\end{split}
\ee 
Therefore, the boundary term purely written in terms of the metric tensor is 
\be 
\begin{split}
\overset{\circ}{\nabla}_{\mu}(Q^{\mu}-\tilde{Q}^{\mu})  & = \dfrac{1}{\sqrt{-g}}  \partial_{\mu}[\sqrt{-g}(g^{\alpha\mu}g^{\beta\nu} - g^{\alpha\nu} g^{\beta\mu} )\partial_{\alpha}g_{\beta\nu}] \\
& = (g^{\mu\rho}g^{\nu\lambda} - g^{\mu\nu}g^{\rho\lambda} ) \partial_{\lambda}\partial_{\rho}g_{\mu\nu} + \dfrac12 (g^{\mu\nu}g^{\rho\gamma}g^{\delta\lambda}-2g^{\mu\rho}g^{\nu\gamma}g^{\delta\lambda} + g^{\mu\gamma}g^{\nu\delta}g^{\rho\lambda} )\partial_{\rho}g_{\mu\nu} \partial_{\lambda}g_{\gamma\delta}.
\label{Bmetric}
\end{split}
\ee 
We observe explicitly that the second order derivatives of the metric are encapsulated purely in this boundary term, as expected. Moreover, it is straightforward to show that the addition of \eqref{Qmetric} and \eqref{Bmetric} give the Ricci scalar, written in terms of the metric tensor and its derivatives
\be 
\overset{\circ}{\mathbb R} = \mathbb Q + ( g^{\mu\rho}g^{\nu\lambda} - g^{\mu\nu}g^{\rho\lambda} ) \partial_{\lambda}\partial_{\rho}g_{\mu\nu} + ( \dfrac12 g^{\mu\gamma}g^{\delta\nu}g^{\rho\lambda} - g^{\rho\mu}g^{\nu\gamma}g^{\delta\lambda} + \dfrac12 g^{\mu\nu}g^{\rho\gamma}g^{\delta\lambda} ) \partial_{\rho}g_{\mu\nu}\partial_{\lambda}g_{\gamma\delta}.
\ee
Yet a last way of writing the nonmetricity scalar comes from the realization that the boundary terms are exactly the terms with partial derivatives on the Levi-Civita connection in the Ricci scalar. Therefore, $\mathbb Q$ corresponds to the $\Gamma \Gamma$ part of $\mathbb R$, that is 
\be
\mathbb Q =  g^{\mu\nu}(\overset{\circ}{\Gamma}{}^{\rho}{}_{\rho\sigma} \overset{\circ}{\Gamma}{}^{\sigma}{}_{\mu\nu} - \overset{\circ}{\Gamma}{}^{\rho}{}_{\mu\sigma}\overset{\circ}{\Gamma}{}^{\sigma}{}_{\rho\nu} ).
\label{QLC}
\ee
Notice that Dirac himself considered the GR Lagrangian in this form, removing the boundary term and considering dependence only on the $\Gamma\Gamma$ part \cite{Dirac:1958sc}. He acknowledges that this choice changes the 3+1 Lagrangian but not the equations of motion. His aim was to obtain the gravitational Hamiltonian, which he obtains together with the primary and secondary constraints of general relativity (denoted in his paper $\phi$ and $\chi$-equations, respectively).

Let us go back to the covariant formalism of symmetric teleparallel gravity, in order to show one last alternative form of writing the nonmetricity scalar in terms of the so-called superpotential
\be 
P^{\rho\mu\nu} = \dfrac{1}{2} Q^{\rho\mu\nu} - \dfrac{1}{2} ( Q^{\mu\nu\rho} + Q^{\nu\mu\rho}) -\dfrac{1}{2} Q^{\rho}g^{\mu\nu} + \dfrac{1}{4}( 2\tilde{Q}^{\rho} g^{\mu\nu} + g^{\rho\mu} Q^{\nu} + g^{\rho\nu}Q^{\mu} ),
\ee 
which is obtained from the variation of the nonmetricity scalar with respect to $Q_{\rho\mu\nu}$, hence its name. With it, the nonmetricity scalar looks like
\be 
\mathbb Q = \dfrac12 P^{\rho\mu\nu}Q_{\rho\mu\nu}.
\ee 
In this form, it is easier to show the equations of motion, which for the metric field can be written as
\be 
\dfrac{2}{\sqrt{-g}}\nabla_{\rho}( \sqrt{-g} P^{\rho}{}_{\mu\nu} ) + \dfrac12 g_{\mu\nu} \mathbb Q + P_{\mu\rho\sigma}Q_{\nu}{}^{\rho\sigma} - 2Q_{\rho\sigma\mu} P^{\rho\sigma}{}_{\nu} = \mathcal{T}_{\mu\nu},
\ee 
where $\mathcal{T}_{\mu\nu}$ is the stress-energy tensor of a matter Lagrangian. Meanwhile the equations of motion for the connection, in the absence of hypermomentum, are \cite{BeltranJimenez:2019tme}
\be 
\nabla_{\mu} \nabla_{\nu} (\sqrt{-g} P^{\mu\nu}{}_{\rho} ) = 0,
\label{connecteqn}
\ee 
are obtained taking the variation with respect to the linear connection. The left-hand side of \eqref{connecteqn} can be proved to vanish identically.

The nonmetricity scalar presented in \eqref{Qstegr} is the particular case of a more general expression for the so-called newer general relativity Lagrangian (in contrast with the new general relativity in the teleparallel formalism), where the nonmetricity scalar is quadratic in five possible combinations of the nonmetricity tensor
\be 
\mathbb Q = c_1 Q_{\rho\mu\nu} Q^{\rho\mu\nu} + c_2 Q_{\rho\mu\nu} Q^{\mu\nu\rho} + c_3 Q_{\mu} Q^{\mu} + c_4 \tilde{Q}_{\mu} \tilde{Q}^{\mu} + c_5 Q_{\mu} \tilde{Q}^{\mu},
\ee 
from which it is obtained the already known results in STEGR by imposing $c_1=\frac{1}{2}$, $c_2=-1$, $c_3=-\frac{1}{2}$, $c_4=0$, and $c_5=1$. Although it could be of interest to perform the 3+1 decomposition of this Newer GR, it would not be reasonable without developing first a solid foundation on the STEGR case, an issue that this article intends to contribute.

\section{The 3+1 split in the STEGR Lagrangian}
\label{sec:split}

In this section we derive the 3+1 split of the STEGR Lagrangian, we compare with previous work on the subject, and propose a different choice of boundary term to compute the Hamiltonian. The choice of boundary term follows the spirit of the teleparallel formalism by avoiding unnecessary second order derivatives of fields in the Lagrangian. We also discuss the effects of different boundary terms, and how these terms are also relevant in GR.

\subsection{ADM decomposition}
Let us first start introducing the traditional ADM split in the metric, by slicing the four-dimensional manifold described by the metric $g_{\mu\nu}$ into three-dimensional hypersurfaces of constant time $\Sigma_t$, equipped with a three-dimensional induced metric $\gamma_{ij}$. We introduce the lapse $\alpha$ and shift $\beta^{i}$ functions, such that the four-dimensional metric is decomposed in the ADM metric form as
\be
g_{00}=-\alpha^2+\beta^i\beta^j\gamma_{ij}, \quad g_{0i}=\beta_i,\quad g_{ij}=\gamma_{ij},
\label{eq:ADMmetric}
\ee
and the inverse components are
\be
g^{00}=-\frac{1}{\alpha^2},\quad g^{0i}=\frac{\beta^i}{\alpha^2},\quad g^{ij}=\gamma^{ij}-\frac{\beta^i\beta^j}{\alpha^2}.
\label{eq:ADMinv}
\ee
This construction is also provided with a normal vector $n_{\mu}$ orthonormal to the induced metric. It satisfies the normalization condition $n_{\mu} n^{\mu}=-1$. Its components are $n_{\mu} =(-\alpha, 0, 0, 0)$ and raising its indices with the inverse ADM metric gives $n^{\mu} = \frac{1}{\alpha} (1, - \beta^{i})$. 

It is convenient to define the so-called projector
\be 
\gamma_{\mu\nu} = g_{\mu\nu} + n_{\mu} n_{\nu}
\ee
which, given the previous ADM decomposition, is nothing more than the induced metric $\gamma_{ij}$. However, this allows to standarize the index notation in order to refer to the induced metric with spacetime indices, even when the temporal components are trivially zero, as it can be easily demonstrated. The projector or induced metric has the role of projecting any tensor component onto the hypersurface. It is also always perpendicular to the normal vector, that is, $n^{\mu}\gamma_{\mu\nu} = n_{\mu}\gamma^{\mu\nu}=0$. 

With these definitions at hand, one possible way of proceeding is to compute the induced decomposition of the full nonmetricity tensor $Q_{\rho\mu\nu}$ in terms of ${n_{\rho}, \gamma_{\mu\nu}}$. The most general decomposition looks like
\be 
\begin{split}
Q_{\alpha\beta\gamma} & = -n_{\alpha}n_{\beta}n_{\gamma}n^{\mu}n^{\nu}n^{\rho} Q_{\mu\nu\rho} + n_{\beta} n_{\gamma} n^{\mu} n^{\nu} Q_{\rho\mu\nu} \gamma_{\alpha}{}^{\rho}  \\
& + n_{\alpha} n_{\gamma} n^{\mu} n^{\nu} Q_{\mu\rho\nu} \gamma_{\beta}{}^{\rho} 
 -  n_{\gamma} n^{\nu} Q_{\nu\rho\mu}  \gamma_{\alpha}{}^{\mu} \gamma_{\beta}{}^{\nu}
+  n_{\alpha} n_{\beta} n^{\mu}n^{\nu} Q_{\mu\nu\rho}  \gamma_{\gamma}{}^{\rho} \\
& -  n_{\beta}n^{\mu} Q_{\nu\mu\rho}  \gamma_{\alpha}{}^{\nu} \gamma_{\gamma}{}^{\rho}
-  n_{\alpha} n^{\mu} Q_{\mu\nu\rho}  \gamma_{\beta}{}^{\nu} \gamma_{\gamma}{}^{\rho}
+ Q_{\mu\nu\rho}  \gamma_{\alpha}{}^{\mu} \gamma_{\beta}{}^{\nu} \gamma_{\gamma}{}^{\rho},
\end{split}
\ee 
and to compute the resulting nonmetricity scalar, as it has been done in \cite{Dambrosio:2020wbi}. However, with this brute force approach we are not taking advantage of the fact that the nonmetricity tensor depends on the metric, for which we already have a 3+1 decomposition, and the scalar fields $\xi^{a}$, whose 3+1 decomposition could be simply regarded as $\xi^{a}=(\xi^{0},\xi^{i})$. We could ask ourselves if such a feat is worth the effort, as we might simply go for the coincident gauge and make this considerations unnecessary. However, one could have the same ruminations regarding the 3+1 decomposition in GR, and fix the lapse and shift in order to not have to deal with their gauge-like but nontrivial behavior. As the simplest 1+1 case shows, the gauge dynamics is nontrivial and can be subject to unpleasant features such as gauge shocks \cite{Alcub}, which suggest that covariant approaches could not be so ingenuous if the gauge fields have ill-posed behavior. Since the connection equation is trivially zero, we would not have to concern about the evolution of the connection in pure STEGR, but only in its nonlinear extensions. We sincerely hope that in the near future these issues will find more interest in the teleparallel community and will be the subject of forthcoming work.

Traditionally, the extrinsic curvature is defined as the rate at which the normal vector to the hypersurface varies, and is given by the formula
\be 
K_{\mu\nu} = -\gamma^{\rho}{}_{\mu} \gamma^{\sigma}{}_{\nu} \overset{\circ}{\nabla}_{\rho} n_{\sigma}.
\label{trueK}
\ee 
This might seem inconsistent, since the connection we are using is not the Levi-Civita, but the symmetric teleparallel one. The previous definition is a geometrical one, but if one insists in defining an analogous extrinsic curvature but using the symmetric teleparallel connection, that is,
\be 
\hat{K}_{\mu\nu} = -\gamma^{\rho}{}_{\mu} \gamma^{\sigma}{}_{\nu} \nabla_{\rho} n_{\sigma},
\label{Kfalse}
\ee 
then the covariant derivative of the normal vector taken with a teleparallel connection, which expressed in terms of the Levi-Civita connection and the distortion tensor will give
\be
\nabla_{\mu} n_{\nu} = \overset{\circ}{\nabla}_{\mu} n_{\nu} - L^{\lambda}{}_{\mu\nu} n_{\lambda} 
\ee 
and consequently,
\be 
\begin{split}
-\gamma^{\rho}{}_{\mu} \gamma^{\sigma}{}_{\nu} \nabla_{\rho} n_{\sigma} & = -\gamma^{\rho}{}_{\mu} \gamma^{\sigma}{}_{\nu} \overset{\circ}{\nabla}_{\rho} n_{\sigma} + \gamma^{\rho}{}_{\mu} \gamma^{\sigma}{}_{\nu} L^{\lambda}{}_{\rho\sigma} n_{\lambda} \\
\hat{K}_{\mu\nu} & = K_{\mu\nu} + \gamma^{\rho}{}_{\mu} \gamma^{\sigma}{}_{\nu} L^{\lambda}{}_{\rho\sigma} n_{\lambda}.
\end{split}
\ee 
Also, some well known properties of the extrinsic curvature such as the fact that it is symmetric in the indices $\mu\nu$ are altered due to the presence of torsion in the connection. Therefore, extra components
\be 
K_{[\mu\nu]} = \gamma^{\rho}{}_{\mu} \gamma^{\sigma}{}_{\nu} n_{\lambda} T^{\lambda}{}_{\rho\sigma}
\ee 
would form part of the kinematic description of the hypersurfaces, however its physical role would be unclear and even problematic if not zero. In TEGR in the Weitzenb\"{o}ck gauge the torsion tensor depends on the (co)frame components $\theta^{a}{}_{\mu}$ and $e_{a}{}^{\mu}$ as
\be 
T^{\rho}{}_{\mu\nu} = e_{a}{}^{\rho}(\partial_{\mu} \theta^{a}{}_{\nu} - \partial_{\nu}\theta^{a}{}_{\mu} ).
\ee 
If we choose the tetrad such that the torsion tensor vanishes, then the antisymmetric part of the extrinsic curvature would also vanish. Unfortunately, these considerations have not been discussed enough in previous research \cite{Capozziello:2021pcg} and will be addressed in future publications.

In \eqref{trueK} we can alternatively write the extrinsic curvature in terms of the 3 dimensional Levi-Civita covariant derivative $D_{\mu}$ univocally defined by the intrinsic metric $\gamma_{\mu\nu}$ as
\be 
K_{\mu\nu} = \dfrac{1}{2\alpha}\left(-\dot{\gamma}_{\mu\nu} + D_{\mu} \beta_{\nu} + D_{\nu} \beta_{\mu} \right).
\ee 
The computation of the 3+1 decomposition of the STEGR action in the coincident gauge can be greatly facilitated by the formula \eqref{QLC}. This is because instead of projecting all the components of the nonmetricity tensor, we can easily compute the ADM decomposition of the Levi-Civita tensor for all components, provided the formulas \eqref{eq:ADMmetric} and \eqref{eq:ADMinv}. These results can also be found in \cite{Alcub}, and correspond to
\begin{align}
\overset{\circ}{\Gamma}{}^{0}{}_{00} & = \dfrac{1}{\alpha}( \partial_t \alpha + \beta^{i} \partial_{i}\alpha - \beta^{i}\beta^{j}K_{ij}), \\
\overset{\circ}{\Gamma}{}^{0}{}_{0i} & = \dfrac{1}{\alpha}(\partial_i \alpha - \beta^{j}K_{ij}), \\
\overset{\circ}{\Gamma}{}^{0}{}_{ij} & = -\dfrac{1}{\alpha} K_{ij}, \\
\overset{\circ}{\Gamma}{}^{i}{}_{00} & = \alpha \partial^{i}\alpha - 2\alpha \beta^{j} K^{i}{}_{j} - \dfrac{1}{\alpha}\beta^{i}(\partial_t \alpha + \beta^{j}\partial_{j}\alpha - \beta^{j}\beta^{k}K_{jk} ), \\
& + \partial_t \beta^{i} + \beta^{j}D_j \beta^{i}, \\
\overset{\circ}{\Gamma}{}^{i}{}_{j0} & = -\dfrac{1}{\alpha}\beta^{i}(\partial_{j}\alpha - \beta^{k}K_{jk} ) - \alpha K^{i}{}_{j} + D_{j}\beta^{i}, \\
\overset{\circ}{\Gamma}{}^{k}{}_{ij} & = \Gamma^{k}{}_{ij} + \dfrac{1}{\alpha} \beta^{k} K_{ij}.
\end{align}
Note that we denote $\Gamma^{k}{}_{ij}$ as the Levi-Civita connection associated to the induced metric $\gamma_{ij}$. We can also compute the contracted components of the Levi-Civita connection $\overset{\circ}{\Gamma}{}^{\alpha}=g^{\mu\nu}\overset{\circ}{\Gamma}{}^{\alpha}{}_{\mu\nu}$, which are
\be 
\overset{\circ}{\Gamma}{}^{0} = -\dfrac{1}{\alpha^3}(\partial_t \alpha - \beta^{i}\partial_{i}\alpha + \alpha^2 K ),
\ee 
\be 
\overset{\circ}{\Gamma}{}^{i} = \Gamma^{i} + \dfrac{\beta^{i}}{\alpha^3}(\partial_t \alpha - \beta^{j}\partial_{j}\alpha + \alpha^2 K ) - \dfrac{1}{\alpha^2}(\partial_t \beta^{i} - \beta^{j}\partial_{j}\beta^{i} + \alpha \partial^{i}\alpha ),
\ee 
with $\Gamma^{i} = \gamma^{jk} \Gamma^{i}{}_{jk}$.

\subsection{3+1 STEGR Lagrangian}

With these expressions it is still lengthy, but straightforward, to compute the 3+1 decomposition of the nonmetricity scalar in its form \eqref{QLC}. Intermediate steps of the computation can be found in several references such as \cite{DAmbrosio:2020nqu}, \cite{Hu:2022anq}, \cite{DAmbrosio:2023asf} and \cite{Tomonari:2023wcs}, in order of appearance. Before performing integrations by parts, this  intermediate step gives
\begin{equation}
\label{mysplit}
\begin{split}
    \mathbb Q & =  
    -{}^{(3)}\mathbb Q - K^{ij} K_{ij} + K^2 
    - \dfrac{1}{\alpha} K \partial_i\beta^{i} - \dfrac{1}{\alpha^3}  \partial_i \beta^{i} \partial_0 \alpha + \dfrac{1}{\alpha^3}( \partial_{i}\alpha  - \alpha \Gamma^{j}{}_{ij} )\partial_0\beta^i  \\
& +\dfrac{1}{\alpha}\left[ \Gamma^{i\hphantom{j}j}_{\hphantom{i}j} - \Gamma^{ji}{}_{j} \right] \partial_{i} \alpha
+ \dfrac{1}{\alpha^3} ( \beta^{i}\partial_j \beta^j - \beta^{j} \partial_{j}\beta^{i}  ) \partial_i \alpha + \dfrac{1}{\alpha^2}( \Gamma^{k}{}_{jk}\beta^{i} \partial_i \beta^{j} + \partial_i\beta^{j} \partial_j \beta^{i} ), \\
\end{split}
\end{equation}
where it has been defined the 3 dimensional nonmetricity scalar as
\be 
{}^{(3)} \mathbb Q = \Gamma^{i}{}_{ij} \Gamma^{jk}{}_{k} - \Gamma^{i}{}_{jk} \Gamma^{j\hphantom{i}k}_{\hphantom{j}i}.
\ee 
Eq. \eqref{mysplit} coincides with Eq.(17) of \cite{DAmbrosio:2020nqu} if taking the coincident gauge $Q_{kij}=\partial_k \gamma_{ij}$. After some tricks and integration by parts of spacetime and spatial derivatives, it is obtained the 3+1 split of the STEGR action as
\be 
 S= -\dfrac{1}{2\kappa} \int d^4x \sqrt{\gamma} \alpha \left[ -{}^{(3)} \mathbb Q - D_i( Q^{i} - \tilde{Q}^{i}) +K^2 - K^{ij}K_{ij}  \right].
 \label{Seh}
\ee 
Without making any further step, this action is equivalent to the Einstein-Hilbert action of GR, since it can be proved that
\be 
{}^{(3)}\mathbb R = -{}^{(3)}\mathbb Q - D_{i}( Q^{i} - \tilde{Q}^{i}).
\ee 
Therefore, the expression \eqref{Seh} is not truly manifesting the modifications to the Lagrangian that are expected to occur once fully considering the nonmetricity scalar, and we can not explore the differences and  potential advantages of the geometrical trinity approach in the 3+1 decomposition. Due to the presence of second order spatial derivatives encapsulated in the term $\alpha D_{i}( Q^{i} - \tilde{Q}^{i})$, we feel motivated to perform an integration by parts and pass the spatial derivative to the lapse function as $\alpha D_{i}( Q^{i} - \tilde{Q}^{i}) = D_i[\alpha( Q^{i} - \tilde{Q}^{i}) ] - D_i \alpha( Q^{i} - \tilde{Q}^{i})$. The term $D_i[\alpha( Q^{i} - \tilde{Q}^{i}) ]$ is a genuine boundary term and is integrated out, and therefore the 3+1 splitted action for STEGR acquires the form
\be 
 S= \dfrac{1}{2\kappa} \int d^4x \mathcal{L} = - \dfrac{1}{2\kappa} \int d^4x \sqrt{\gamma} \alpha \left[ K^2 - K^{ij}K_{ij} -{}^{(3)} \mathbb Q + \dfrac{\partial_i \alpha}{\alpha}(Q^{i} - \tilde{Q}^{i}) \right],
 \label{Strue}
\ee 
which now does not contain second order derivatives on any fundamental variable. Although one could start this integration by parts procedure from the 3+1 Lagrangian for general relativity without any reference of STEGR, in our opinion there would be no motivation for that without the notion of the 3D nonmetricity scalar and the boundary term. In this sense, \eqref{Strue} resembles more a genuine 3+1 decomposition for STEGR due to the lack of second order derivatives on the metric, although many other combinations could be possible from the terms that were integrated by parts and were not included in this final result. Whether some of them are useful for numerical relativity purposes, will be tested in future investigations.

\subsection{The Hamiltonian and the effect of different boundary terms}

The new action \eqref{Strue} for STEGR features a slightly different Lagrangian which will change the canonical Hamiltonian for STEGR, as we will show in this subsection. First, the definition of the canonical momenta does not change regarding its GR counterpart, since
\be 
\pi^{ij} = \dfrac{\partial \mathcal{L}}{\partial \dot{\gamma}_{ij}} = \sqrt{\gamma} ( K^{ij} - \gamma^{ij}K ).
\label{picurv}
\ee 
With this we compute the expression for the STEGR canonical Hamiltonian
\be 
\mathcal{H} = \dot{\alpha} \pi + \dot{\beta}_i \pi^{i} + \dot{\gamma}_{ij}\pi^{ij} - \mathcal{L}.
\ee 
In the previous equation we are considering lapse and shift as variables belonging to the configuration space, and $\pi$ and $\pi^{i}$ their respective conjugate momenta. However, they are constrained to be zero since the Lagrangian does not contain time derivatives of $\alpha$ and $\beta_i$. This means that
\begin{equation}
    \pi = \dfrac{\partial \mathcal{L}}{\partial \dot{\alpha}} = 0, \qquad \pi^{i} = \dfrac{\partial \mathcal{L}}{\partial \dot{\beta}_{i}} = 0,
\end{equation}
which act as primary constraints and will be later added to the primary Hamiltonian multiplied by Lagrange multipliers $\lambda$ and $\lambda_i$, respectively. Then, we obtain that the canonical Hamiltonian for STEGR in the coincident gauge is 
\begin{equation}
\mathcal{H} = \dfrac{\alpha}{\sqrt{\gamma}}\left( \pi_{ij}\pi^{ij} - \dfrac12 \pi^2 \right) + \alpha \sqrt{\gamma} {}^{(3)} \mathbb Q - \sqrt{\gamma} \partial_i \alpha(Q^{i}-\tilde{Q}^{i} )  - 2 \sqrt{\gamma} \beta_i D_j (\pi^{ij}/\sqrt{\gamma}).
\label{NewHamilt}
\end{equation} 
Here, we realize that if we want to write the canonical Hamiltonian as a sum of Hamiltonian and momenta constraints as in GR, then the presence of the spatial derivative of $\alpha$ in the expression is an obstruction to such interpretation as a Lagrange multiplier. If we insist in it, then the canonical Hamiltonian is written as
\begin{equation}
  \mathcal{H}   = \alpha \mathcal{H}_0 + \beta_{i} \mathcal{H}^i,
\end{equation}
which is the sum of the proposed Hamiltonian $\mathcal{H}_0$ and momenta $\mathcal{H}_i$ constraints, with lapse function $\alpha$ and shift vector $\beta^{i}$ as Lagrange multipliers. The new Hamiltonian and momenta constraints are, respectively
\be 
\begin{split}
 \mathcal{H}_0 & = \dfrac{1}{\sqrt{\gamma}}\left( \pi^{ij}\pi_{ij}-\dfrac12 \pi^2 \right) + \sqrt{\gamma}{}^{(3)}\mathbb Q - \sqrt{\gamma}\dfrac{\partial_i \alpha}{\alpha} (Q^{i} - \tilde{Q}^{i} ),\\
 \mathcal{H}^i & = -2 \sqrt{\gamma} D_j ( \pi^{ij} /\sqrt{\gamma} ).
\end{split}
\ee 
As we will discuss later, a similar situation arises in the TEGR Hamiltonian, where  the Hamiltonian constraint is analogously modified in the terms that do not depend on the extrinsic curvature. These changes are expected, given to the well-known fact that the TEGR, STEGR and GR Lagrangians differ from each other by boundary terms. In the next section, it will be more convenient to express the STEGR Hamiltonian constraint in terms of the extrinsic curvature as
\be 
\sqrt{\gamma} \left[K_{ij}K^{ij} - K^2 + {}^{(3)}\mathbb Q    - \dfrac{\partial_i \alpha}{\alpha} (Q^{i} - \tilde{Q}^{i} )\right] = 0,
\label{HsymLagr}
\ee 
since the canonical momenta and the extrinsic curvature can be used interchangeably due to Eq. \eqref{picurv}.

\subsection{Boundary terms}

In GR, the Einstein-Hilbert action requires the addition of boundary terms in order to obtain a well-posed variational principle on manifolds with boundaries, including asymptotic boundaries. This is the well-known Gibbons-Hawking-York boundary term, although it has been shown to not be unique, as new variables can be introduced that keep the variational principle well-posed \cite{Chakraborty:2016yna}. This reference also discusses the argument that the ADM Lagrangian contains no time derivatives of lapse or shift. However, adding boundary terms to the ADM Lagrangian leads to a new Lagrangian which contain time derivatives of them. Nevertheless, the lapse and shift remain non-dynamical, and their conjugate momenta cannot be inverted. 

The fact that the Hamiltonian constraint for STEGR is changed due to the choice \eqref{Strue}, which affects the 3+1 Lagrangian, should not be surprising. Any alternative choice of boundary terms in the Lagrangian will inevitably alter the canonical momenta, the Hamiltonian, and other canonical structures sensitive to boundary terms. It is worth noting that we have implemented the integration by parts in the Lagrangian only after applying the 3+1 decomposition of the metric. This decomposition is not unique, since there are infinite ways of carrying out this integration. 

For STEGR, it is established that the Ricci scalar and the nonmetricity scalar are equivalent up to a boundary term. Similarly, in TEGR the Ricci scalar differs from the torsion scalar by another boundary term. Based in this fact, it has been suggested in the literature that this TEGR boundary term might play the role of the Gibbons-Hawking-York boundary term \cite{Oshita:2017nhn}, which can be related to the black hole entropy by standard thermodynamic arguments. However, it has been shown that this identification of the two boundary terms is incomplete, and the equivalence depends on the choice of frame or tetrad, up to local Lorentz transformations that determine a specific given metric \cite{Fiorini:2023axr}. In the latter reference, the GHY term is formulated in the context of torsion, which leads to the right black hole entropy without needing to perform a background subtraction. The frames constructed there have been adapted for a Schwarzschild spacetime, and also yield the correct value of the gravitational energy calculated from the energy-momentum pseudo-current in the teleparallel framework.

\section{Hamilton's equations in STEGR}
\label{sec:Hamilteqns}

In order to illustrate the impact that the choice of boundary term produces in a concrete way, we compute Hamilton's equations for the modified STEGR Hamiltonian proposed in \eqref{NewHamilt}. For this purpose, we need the primary Hamiltonian by incorporating the primary constraints previously identified. Then, the primary Hamiltonian reads
\begin{align}
    \mathcal{H}_{p} & = \mathcal{H} + \lambda \pi + \lambda_i \pi^{i} \\
    & = \dfrac{\alpha}{\sqrt{\gamma}}\left( \pi_{ij}\pi^{ij} - \dfrac12 \pi^2 \right) + \alpha \sqrt{\gamma} {}^{(3)} \mathbb Q - \sqrt{\gamma} \partial_i \alpha(Q^{i}-\tilde{Q}^{i} )  - 2\beta_i D_j \pi^{ij} + \lambda \pi + \lambda_i \pi^{i} .
\end{align}
It is important to note that the additional term in the Hamiltonian that arises from integration by parts is independent of the canonical momenta $\pi, \pi^{i}, \pi^{ij}$. Consequently, the following Hamilton's equations remain unchanged relative to their GR counterparts 
\begin{equation}
    \dot{\alpha} = \dfrac{\partial \mathcal{H}}{\partial \pi} = \lambda 
\end{equation}
\begin{equation}
    \dot{\beta_i} = \dfrac{\partial \mathcal{H}}{\partial \pi^{i}} = \lambda_i
\end{equation}
\begin{equation}
    \dot{\gamma_{ij}} = \dfrac{\partial \mathcal{H}}{\partial \pi^{ij}} = \dfrac{2}{\sqrt{\gamma}} \alpha \left( \pi_{ij} - \dfrac12 \gamma_{ij} \pi \right) + 2 D_{(i}
\beta_{j)}.
\end{equation}
To compute the Hamilton equation for $\dot{\pi}$, we notice a difference regarding the GR case in the term depending on $\partial_i \alpha$. However, the spatial derivative can be passed to the factor $\sqrt{\gamma}(Q^{i} - \tilde{Q}^{i} )$, and we are able to perform the variation in the same as in GR, obtaining 
\begin{equation}
    \dot{\pi} = -\dfrac{\partial \mathcal{H}}{\partial \alpha} = -\dfrac{1}{\sqrt{\gamma}}\left( \pi^{ij}\pi_{ij} - \dfrac12 \pi^2 \right) - \sqrt{\gamma} {}^{(3)}\mathbb Q - \partial_i [\sqrt{\gamma} (Q^{i}-\tilde{Q}^{i}) ]
\end{equation}
In addition, the equation for $\dot{\pi}^{i}$ is unchanged
\begin{equation}
    \dot{\pi}^{i} = - \dfrac{\partial \mathcal{H}}{\partial \beta_i} = -2 \sqrt{\gamma} D_j (\pi^{ij} /\sqrt{\gamma} ).
\end{equation}
Remarkably, the Hamilton equations for $\dot{\pi}^{ij}$ are modified due to the new dependence of the primary Hamiltonian on the intrinsic metric $\gamma_{ij}$. The computation is more involved than in previous equations, so we carry it out in several steps. We begin by computing the variation of all terms with respect to the metric, which yields
\begin{equation}
\begin{split}
  \dot{\pi}^{ab} & = - \dfrac{\delta H}{\delta \gamma_{ab}} =  - \alpha \dfrac{\delta}{\delta \gamma_{ab}} \left(\dfrac{1}{\sqrt{\gamma}} \right)(\pi_{ij} \pi^{ij} - \dfrac{1}{2} \pi^2 ) - \dfrac{\alpha}{\sqrt{\gamma}} \dfrac{\delta}{\delta \gamma_{ab}}\left( \pi_{ij}\pi^{ij} - \dfrac{1}{2} \pi^2 \right) - \dfrac{\delta \sqrt{\gamma}}{\delta \gamma_{ab}} \left[ \alpha {}^{(3)}\mathbb Q - \partial_i \alpha \left( Q^{i} - \tilde{Q}^{i} \right) \right] \\
    & -\sqrt{\gamma} \alpha \dfrac{\delta {}^{(3)}\mathbb Q}{\delta \gamma_{ab}} - \sqrt{\gamma} \partial_i \alpha \dfrac{\delta}{\delta \gamma_{ab}}(Q^{i} - \tilde{Q}^{i} ).
    \label{rawHeqpi}
\end{split}
\end{equation}
The last line involves the variation of the spatial nonmetricity scalar ${}^{(3)}\mathbb Q$, which gives the following
\begin{equation}
\begin{split}
{}^{(3)}Q^{ab} & \equiv \dfrac{\delta {}^{(3)} \mathbb Q}{\delta \gamma_{ab}}  = -\gamma^{ai} \gamma^{bj} (\Gamma^{k}{}_{ij} \Gamma^{l}{}_{kl} - \Gamma^{k}{}_{jl} \Gamma^{l}{}_{ik} ) \\
& -\dfrac{1}{2}\gamma^{ij}\Gamma^{l}{}_{kl} \left( \gamma^{ka}  \Gamma^{b}{}_{ij} + \gamma^{kb}   \Gamma^{a}{}_{ij}  \right) + \Gamma^{l}{}_{kl}( -\partial_i\gamma^{kb} \gamma^{ai} - \partial_i \gamma^{ka} \gamma^{ib} + \partial_i \gamma^{ki} \gamma^{ab} )\\
&-\dfrac{1}{2}\gamma^{ij} \Gamma^{k}{}_{ij} \left( \gamma^{la} \Gamma^{b}{}_{kl}  + \gamma^{lb}  \Gamma^{a}{}_{kl}  \right)  - \partial_{k} \gamma^{ab} \gamma^{ij} \Gamma^{k}{}_{ij} + \gamma^{ij} \Gamma^{l}{}_{jk} ( \gamma^{ka} \Gamma^{b}{}_{il} + \gamma^{kb} \Gamma^{a}{}_{il} ) \\
&  + \gamma^{ij} ( \partial_{i} \gamma^{kb} \Gamma^{a}{}_{jk} + \partial_{i} \gamma^{ka} \Gamma^{b}{}_{jk} ) + \Gamma^{i}{}_{jk}( \partial_{i}\gamma^{kb} \gamma^{aj} + \partial_{i}\gamma^{ka} \gamma^{bj} ) -  \partial_{i} \gamma^{ki} ( \gamma^{aj} \Gamma^{b}{}_{jk} + \gamma^{bj} \Gamma^{a}{}_{jk} ).
\end{split}
\end{equation}
Meanwhile, the variation of the term involving the two traces of the nonmetricity tensor with respect to the intrinsic metric gives
\begin{equation}
\begin{split}
{}^{(t)}Q^{ab} & \equiv \partial_i \alpha \dfrac{\delta}{\delta \gamma_{ab}} (Q^{i}-\tilde{Q}^i) = \dfrac{\partial_i \alpha}{2} \left( - \gamma^{bi}(Q^{a} - \tilde{Q}^{a}) - \gamma^{ai}( Q^{b} - \tilde{Q}^{b} )  \right) \\
&
 + \dfrac{\partial_i \alpha}{2} \left( -\partial_{c} \gamma_{dj} \gamma^{ci}\left[ \gamma^{ad} \gamma^{bj} + \gamma^{aj} \gamma^{bd} \right] + \partial_{c}\gamma_{dj} \gamma^{di} \left[ \gamma^{ac} \gamma^{bj}+ \gamma^{aj} \gamma^{bc} \right] \right).
 \end{split}
\end{equation}
It is important to keep in mind that these variations are only valid in the coincident gauge case, where the nonmetricity tensor is simply the partial derivative of the metric. In the most general case, there will be more intricate dependencies of the connection. Replacing everything back in \eqref{rawHeqpi} and computing the standard variations in $\sqrt{\gamma}$ and the $\pi^2$ terms, we obtain
\begin{equation}
\begin{split}
    \dot{\pi}^{ij} &  = - \dfrac{\partial \mathcal{H}}{\partial \gamma_{ij}} = \dfrac{1}{2\sqrt{\gamma}} \gamma^{ij} \alpha (\pi_{kl} \pi^{kl} - \dfrac12 \pi^2 ) - \dfrac{1}{\sqrt{\gamma}} \alpha ( 2\pi^{ik}\pi^{j}{}_{k} - \pi^{ij} \pi ) -\dfrac12 \sqrt{\gamma} \gamma_{ij} \left[  \alpha Q - \partial_i \alpha(Q^{i} - \tilde{Q}^{i} ) \right] \\
    & - \sqrt{\gamma} \alpha {}^{(3)}Q^{ij} - \sqrt{\gamma} {}^{(t)}Q^{ij} ,
    \end{split}
\end{equation}
We can compare this evolution equation for the momenta with its GR counterpart
\begin{equation}
    \begin{split}
        \dot{\pi}^{ij} = & -\dfrac{\delta H}{\delta \gamma_{ij} } = -\alpha \sqrt{\gamma}\left( {}^{(3)}R^{ij} - \dfrac12 {}^{(3)}R \gamma^{ij} \right) + \dfrac{1}{2 \sqrt{\gamma}} \alpha  \gamma^{ij} \left( \pi^{kl}\pi_{kl} - \dfrac12 \pi^2 \right) \\
        &- \dfrac{1}{\sqrt{\gamma}} \alpha  \left( 2\pi^{ik}\pi_{k}{}^{j} -  \pi \pi^{ij} \right)   + \sqrt{\gamma} (D^{i}D^{j}\alpha - \gamma^{ij} D^{k}D_{k}\alpha ) + \sqrt{\gamma} D_k(  \beta^k \pi^{ij}/\sqrt{\gamma} ) \\
        &- 2\pi^{k(i} D_{k}\beta^{j)},
    \end{split}
\end{equation}
as it originally appears in Eq.(3.15b) in \cite{Arnowitt:1962hi} or in Eq.(100) in \cite{Blixt:2023kyr}. One of the main differences is the absence of second-order spatial derivatives for the lapse and of first-order derivatives for the shift vector. Instead, we obtain terms with first-order spatial derivatives for the lapse multiplied by first-order spatial derivatives of the intrinsic metric. These new features might present advantages when obtaining a system of differential equations of first-order suitable for hyperbolicity analysis. Since several fields already appear with first-order derivatives, the necessity of auxiliary variables to rewrite the system with only first-order derivatives decreases.

Our point is that a change in the boundary term in the 3+1 Lagrangian does not alter the kinetic terms that provide the equivalence between STEGR and GR, but they change the evolution of the gauge variables $\alpha$ and $\beta_i$. The importance of the evolution of these variables and the fixing of those is exemplified in a classical example with spherical symmetry in  \cite{Alcub}. There it happens that after finding that the ADM equations in spherical symmetry are strongly hyperbolic, there is one exception when the parameter $f(\alpha)$ in the Bona-Masso slicing condition
\begin{equation}
    \partial_t \alpha = - \alpha^{2} f(\alpha) K
\end{equation}
is equal to $1$. In this case, the hyperbolicity fails since some of the so-called eigenfields of the principal symbol of the set of equations are ill-defined. In this case it is necessary to modify the evolution equations, either by using the BSSNOK approach, or by a change of variables. This illustrates that exploring different formulations of the ADM system of equations is worthwhile and can have serious consequences in the appearance of hyperbolicity.

The implications of our modified equation for $\dot{\pi}^{ij}$ can be realistically tested in problems of physical importance, such as spherical symmetry. In the next section, however, we will study the Hamiltonian constraint in spherical symmetry, and the analysis of the evolution equations together with hyperbolicity will be left for future work \cite{GuzmanJaarma2026}. As far as we are aware, this is the first comprehensive explanation of how the time evolution for STEGR can be formulated in such a way that produces testable changes in numerical codes. Notice that it is not necessary to move to the Hamiltonian picture, and this analysis is also possible in the Lagrangian one, via the modified Lagrangian in Eq. \eqref{Strue}. Although the ADM equations are originally formulated in the Hamiltonian formulation \cite{Arnowitt:1962hi}, we expect to work in the future in the Lagrangian formulation \cite{GuzmanJaarma2026}.

\section{The Hamiltonian constraint for spherical symmetry}
\label{sec:example}
In order to illustrate the difference produced at the level of the Hamiltonian constraint in STEGR, let us consider a simple case of a spherically symmetric spacetime with only radial dynamics. The general form for the spatial metric will be, in spherical coordinates $x^{i} = (r, \theta, \varphi)$ \cite{Alcub}
\be 
dl^2 = A(t,r)dr^2 + r^2 B(t,r)\left[d\theta^2 + \sin^2(\theta)d\varphi^2 \right] = \gamma_{ij} dx^{i} dx^{j},
\ee 
with $A$ and $B$ positive metric functions. A short digression here is that this form of the metric is not compatible with the coincident gauge for modifications to the STEGR gravity, see for instance \cite{Bahamonde:2022esv} for the form of the connection compatible with a spherically symmetric spacetime. However, in our case we are only interested in STEGR, for which the connection does not play any role in the equations of motion.

We compute the 3 dimensional Christoffel symbols
\begin{align}
    \Gamma^{i}{}_{jk} = \gamma^{il}( \partial_{j}\gamma_{kl} + \partial_{k}\gamma_{jl} - \partial_{l}\gamma_{jk}  ),
\end{align}
where only the following terms are nonvanishing
\begin{align}
    &  \Gamma^{r}{}_{rr} = \dfrac{\partial_r A}{2A} = \dfrac12 D_A, \\ 
    &  \Gamma^{r}{}_{\theta\theta} = -\dfrac{r}{2A}\left( 2B - r \partial_r B \right) = -\dfrac{rB}{2A}(2 + rD_B), \\
    &  \Gamma^{r}{}_{\varphi\varphi} = -\dfrac{ r \sin^2(\theta)}{2A}\left( 2B - r \partial_r B \right) = -\dfrac{rB \sin^2(\theta)}{2A} (2 + rD_B), \\
    &  \Gamma^{\theta}{}_{r\theta} =  {}^{(3)} \Gamma^{\theta}{}_{\theta r} = \dfrac{1}{2B r}\left( 2B + r \partial_r B \right) = \dfrac{1}{r}(2+rD_B), \\
    &  \Gamma^{\theta}{}_{\varphi\varphi} = -\cos(\theta) \sin(\theta), \\
    &  \Gamma^{\varphi}{}_{r\varphi} = \cot(\theta), \\
    & \Gamma^{\varphi}{}_{r\varphi} = {}^{(3)} \Gamma^{\varphi}{}_{\varphi r} =\dfrac{1}{2Br} \left( 2B  + r\partial_r B \right) = \dfrac{1}{2r}(2+rD_B).
\end{align}
In order to match the results in \cite{Alcub}, we have alternatively expressed the terms of the connection by using the following auxiliary quantities
\begin{align}
    D_{A} \coloneqq \partial_r \ln A, \qquad \ D_B \coloneqq \partial_r \ln B,
\end{align}
which are intended to deal with Einstein equations in first order form. Note that in this setup the extrinsic curvature has only two independent components that will be defined as
\be 
K_A = K^{r}_{r}, \qquad K_B = K^{\theta}_{\theta} = K^{\varphi}_{\varphi}.
\ee 
With this at hand, and as an intermediate step, we show the value of ${}^{(3)}\mathbb Q$ in the original and the auxiliary variables, which gives 
\be 
{}^{(3)} \mathbb Q = \dfrac{1}{2AB^2 r^2} ( 2B+r \partial_r B )^2 = \dfrac{2}{A r^2} + \dfrac{2 D_B}{A r} + \dfrac{(D_B)^2}{2A}.
\ee 
Meanwhile, the expression for the term proportional to $\partial_i \alpha$ in \eqref{HsymLagr} is
\begin{equation}
\dfrac{\partial_i \alpha}{\alpha}(Q^{i}-\tilde{Q}^{i} ) =  \dfrac{4\partial_r \alpha}{rA\alpha} + \dfrac{2 \partial_r B \partial_r \alpha}{AB\alpha} = \dfrac{4 D_{\alpha}}{rA} + \dfrac{2 D_B D_{\alpha}}{A},
\end{equation}
where we have also defined $D_{\alpha} = \partial_r \text{ln}\alpha = \partial_r \alpha/\alpha$. All together, we obtain the formula for the Hamiltonian constraint in STEGR as
\begin{equation}
\begin{split}
\mathcal H_0 & = A K_B ( 2K_A + K_B )  +  \\
& = A K_B ( 2K_A + K_B )  +  \dfrac{2}{A r^2} + \dfrac{2 D_B}{A r} + \dfrac{(D_B)^2}{2A} + \dfrac{4 D_{\alpha}}{rA} + \dfrac{2 D_B D_{\alpha}}{A}
\end{split}
\end{equation}
that is modestly simpler than the general relativity Hamiltonian constraint
\be 
\mathcal H = AK_B(2K_A+K_B)+\dfrac{1}{r^2B}(A-B) - \partial_r D_B + \dfrac{1}{r}(D_A-3D_B) + \dfrac{D_A D_B}{2} - \dfrac{3D_B^2}{4}.
\ee 
We see that the main differences occur at the level of the extra terms containing spatial derivatives in either the functions of the metric $A,B$ or in the lapse $\alpha$, however the terms containing the time derivatives on the intrinsic metric, that is the terms depending on the extrinsic curvature, remain unaltered. Notice that the term with second order derivatives $-\partial_r D_B$ disappears but  instead in exchange we have obtained a term with first order derivatives on the lapse. Although the full consequences of this new form of the constraint remain to be investigated, some preliminary results could indicate that it simplifies the numerical setup of simple work horse example in numerical relativity such as the self interacting scalar field in a spherically symmetric spacetime \cite{GuzmanIntroNR}, potentially having consequences for more complex, three-dimensional examples.\\

Since we have altered the 3+1 GR Lagrangian by a boundary term, the change in the Hamiltonian constraint could be expected, as also the ADM evolution equations in first order form (for this spherically symmetric metric they can be found in the classical literature in numerical relativity \cite{Alcub,Baumgarte:2010ndz,Baumgarte:2021skc}). Although the main difference between STEGR and GR occurs at the level of the Hamiltonian constraint and in the Hamilton equation for $\dot{\pi}^{ij}$ and it can be expressed analytically for problems of astrophysical interest such as spherical symmetry, it would be interesting to analyze if at the numerical level some differences in computation time could be obtained. In addition, it remains to be investigated the consequences for hyperbolicity produced by the different dependence of the lapse and shift in the dynamical equations for the momenta. 

\section{Discussion}
\label{sec:ext}
There are a couple of important points that we would like to address, since they are relevant ways of extending our work for future research with applications to numerical relativity. 
\begin{itemize}
    \item Naturally, the same approach taken here for STEGR can be addressed in TEGR, since the Hamiltonian, therefore the Hamiltonian constraint, are also modified concerning its GR counterpart. However, there are additional subtleties concerning the choice of the orientation of the tetrad, that is, we have freedom to Lorentz rotate it, therefore the issue is more complicated as portrayed here. It is known that the canonical Hamiltonian for TEGR is the addition of the Hamiltonian $\mathcal H_0$ and momenta $\mathcal H_i$ constraints, that is
\be 
\mathcal{H}_\mathrm{TEGR} = \alpha \mathcal{H}_0 + \beta^{i} \mathcal{H}_i
\ee
where the expressions for both constraints are
\be 
\begin{split}
\mathcal{H}_0 &= \frac{ \kappa}{4\sqrt{\gamma}}\left[\Pi_a{}^i\Pi_b{}^l\theta^a{}_k \theta^b{}_j\gamma^{jk}\gamma_{li}  +\Pi_a{}^i\Pi_b{}^j\theta^a{}_j\theta^b{}_i -\Pi_a{}^i\Pi_b{}^j\theta^a{}_i\theta^b{}_j \right]   \\
    &  -\frac{\sqrt{\gamma}}{2\kappa}{}^{(3)} \mathbb{T} - n^a\partial_i \Pi_a{}^i ,
\end{split}
\ee 
and
\be 
\mathcal{H}_i =  -\theta^A{}_j \partial_i \Pi_A{}^i-\Pi_A{}^i T^A{}_{ij}.
\ee
Here we are following the notation in \cite{Pati:2022nwi}. Basically, $\Pi^{i}_a$
 is the conjugate momenta to the tetrad $\theta^{a}_{i}$, and $n^{a}$ is the normal vector with  tangent space indices. We observe that $H_0$ contains terms analogous to the $\pi^2$ terms in the GR Hamiltonian, but also ${}^{(3)}\mathbb T$, which is the analog of the ${}^{(3)}\mathbb Q$ introduced in the context of STEGR. Again, this is equivalent to the Hamiltonian for GR in the dynamical part of the canonical momenta/extrinsic curvature, but the extra nondynamical terms are modified due to the departure of TEGR with respect to GR due to a boundary term. The exploration of the new features appearing in TEGR and their consequences in numerical relativity are currently under study. 
    \item The covariant approaches of TEGR and STEGR offer an additional set of fields encoded in the connection, which can be either the Lorentz matrices $\Lambda^{a}{}_{b}$ or the diffeomorphism fields $\xi^{a}$. It has already been shown in the Hamiltonian picture that the introduction of the $\Lambda$'s increase the number of Hamilton's equations but they are also involved in the 3+1 equations for the tetrad. \cite{Pati:2022nwi}. We expect that analogous features will appear in a full covariant version of STEGR. However, its canonical implementation  is more involved than the TEGR one due to the presence of second order derivatives in the fields $\xi^{a}$ in the STEGR Lagrangian, as it has recently been discussed in \cite{Blixt:2023kyr}. 
\end{itemize}

\section{Conclusions}
\label{sec:concl}
We have presented the mathematical formalism pertinent to the symmetric teleparallel equivalent of general relativity, particularly stressing the similarities and differences with general relativity. Since the GR and STEGR Lagrangians differ by a boundary term, differences in the equations of motion would not be expected. However, we have altered the boundary term via an integration by parts, obtaining a slightly different Lagrangian with altered nondynamical terms. Such boundary terms that relate GR and STEGR affect the canonical structure of any physical theory, a fact that has consequences in other aspects of physics different than the dynamics of the equations of motion, as for instance physics that depends on the boundary such as black hole thermodynamics or the definition of energy.

We have reviewed the 3+1 decomposition of the Lagrangian of STEGR by introducing an ADM split in the metric and derived the Hamiltonian, the Hamiltonian and momenta constraints, and Hamilton's equations for the momenta. Different from previous works, we have presented arguments in favor of altering a boundary term in the action that otherwise would give the same Hamiltonian than general relativity, and used to our favor to obtain a modified 3+1 Lagrangian and therefore a different Hamiltonian constraint. The modification of the boundary term gives outcomes that can be interpreted as if we were exploiting the gauge invariance of GR, that is always present in the choice of the lapse and shift functions. In this way, in our Lagrangian the boundary term has been integrated by parts in order to get rid of second order derivatives, and it has a nature closer to the main spirit of the teleparallel framework, which is to obtain a Lagrangian with only first order derivatives on the metric tensor (or the tetrad, in the case of metric teleparallel gravity). 

One of the conclusions from our considerations is that the Hamiltonian, the Hamiltonian constraint, and the Hamilton equations for $\pi^{ij}$ in STEGR are different than in the general relativity case. This difference is presented explicitly for a simple case of a spherically symmetric spacetime. The expression for the Hamiltonian constraint is simpler, which proves analytically that STEGR could present advantages for numerical relativity. A numerical proof of this claim, the consequences of our approach for hyperbolicity in spherical symmetry, and applications to modified gravity based on the symmetric teleparallel framework are left for future work.\\

{\bf Acknowledgments.} M. J. G. is grateful to Miguel Bezares, Daniel Blixt and Laur Järv for helpful discussions and comments, to Alexey Golovnev and Tomi Koivisto for their comments in the first version of the manuscript, to Laxmipriya Pati for corrections to the second version of the manuscript, and to Marie Femke Jaarma for her encouragement during the last stages of this work. The computations in this paper have been partially assisted or checked with Cadabra 2.0 \cite{{Peeters:2006kp,Peeters:2007wn,Peeters:2018dyg}}. M. J. G. has been supported by the Estonian Research Council grant PSG910 ``Theoretical frameworks for numerical modified gravity''.

\end{document}